# Integrating Machine Learning with Multimodal Monitoring System Utilizing Acoustic and Vision Sensing to Evaluate Geometric Variations in Laser Directed Energy Deposition

Ke Xu, Chaitanya Krishna Prasad Vallabh, Souran Manoochehri[*]

**Department of Mechanical Engineering,**
**Stevens Institute of Technology,**
**Hoboken, NJ 07030, USA**

**Abstract**

Laser directed energy deposition (DED) additive manufacturing struggles with consistent part quality due to complex melt pool dynamics and process variations. While much research targets defect detection, little work has validated process monitoring systems for evaluating melt pool dynamics and process quality. This study presents a novel multimodal monitoring framework, synergistically integrating contact-based acoustic emission (AE) sensing with coaxial camera vision to enable layer-wise identification and evaluation of geometric variations in DED parts. The experimental study used three part configurations: a baseline part without holes, a part with a 3mm diameter through-hole, and one with a 5mm through-hole to test the system's discerning capabilities. Raw sensor data was preprocessed: acoustic signals were filtered for time-domain and frequency-domain feature extraction, while camera data underwent melt pool segmentation and morphological feature extraction. Multiple machine learning algorithms (including SVM, random forest, and XGBoost) were evaluated to find the optimal model for classifying layer-wise geometric variations. The integrated multimodal strategy achieved a superior classification performance of 94.4%, compared to 87.8% for AE only and 86.7% for the camera only. Validation confirmed the integrated system effectively captures both structural vibration signatures and surface morphological changes tied to the geometric variations. While this study focuses on specific geometries, the demonstrated capability to discriminate between features establishes a technical foundation for future applications in characterizing part variations like geometric inaccuracies and manufacturing-induced defects.

**Keywords:** Laser Directed Energy Deposition, Multimodal Monitoring, Acoustic Emission, Machine Learning, Geometric Variation Classification, Additive Manufacturing, Quality Control

---

[*] Corresponding author.
 E-mail address: smanooch@stevens.edu (S. Manoochehri).

# 1. Introduction

  Laser Directed Energy Deposition (DED) has emerged as a prominent additive manufacturing (AM) technology, particularly valued for its capability to fabricate complex geometries, repair existing components, and produce functionally graded materials [1,2]. In DED processes, a high-energy laser beam selectively melts and fuses metallic powders or wires delivered to the substrate surface, enabling layer-by-layer construction of three-dimensional parts with considerable design freedom [3]. This technology has gained significant traction in aerospace, automotive, and energy industries, where high-performance components with complex internal structures and enhanced mechanical properties are essential [4]. The versatility of DED extends beyond conventional manufacturing limitations, offering notable advantages including the ability to process a wide range of materials, achieve high deposition rates, and enable multi-material fabrication within a single build [5]. Moreover, DED provides greater flexibility in part size constraints and allows for the addition of material to existing structures, making it particularly suitable for repair and remanufacturing applications [6]. Despite these advantages, DED processes face substantial challenges related to process reliability and part quality consistency due to complex thermal dynamics, rapid solidification rates, and intricate laser-material interactions that can lead to various defects compromising mechanical integrity and dimensional accuracy [7].

  The challenges in DED manufacturing originate from the inherently complex nature of the process itself. These process complexities make DED processes particularly susceptible to various defects that can significantly compromise the mechanical properties and dimensional accuracy of fabricated components. The complex interplay between laser parameters, powder characteristics, and thermal dynamics creates conditions conducive to defect formation, including porosity, cracking, geometric irregularities, and microstructural inconsistencies [8,9]. These defects arise from the rapid heating and cooling cycles characteristic of DED, which can lead to residual stress accumulation, incomplete fusion between layers, and non-uniform material properties throughout the build [10]. Among the commonly observed defects, porosity represents a critical concern due to its negative effects on mechanical strength and fatigue resistance, resulting from gas entrapment during rapid solidification, incomplete powder melting, and keyhole formation [11]. Similarly, cracking defects, including hot tears and solidification cracks, frequently occur due to thermal stress concentrations and rapid cooling rates, particularly in materials with high thermal expansion coefficients [12]. The underlying physics governing defect formation involves complex heat transfer mechanisms, fluid dynamics in the melt pool, and solidification kinetics that are difficult to predict and control [13]. Furthermore, thermal gradients generated during the process create residual stresses that can exceed material yield strength, while melt pool dynamics, influenced by laser energy input and powder characteristics, determine the extent of material fusion and the likelihood of defect incorporation [14].

  Recognizing the severity of these quality challenges, their impact extends beyond individual part performance to broader manufacturing considerations, including increased post-processing requirements, higher rejection rates, and reduced process reliability [15]. Variability in mechanical properties due to defect presence can compromise the structural integrity of critical components, as demonstrated in aerospace applications [4]. Consequently, the development of effective monitoring and quality control strategies has become essential for enabling wider industrial adoption of DED technology and ensuring consistent production of high-quality components.

In response to these quality assurance needs, researchers have explored various monitoring approaches, with vision-based systems emerging as one of the widely adopted strategies. Vision-based monitoring systems have gained considerable attention due to their ability to provide direct visual feedback on melt pool characteristics and layer formation quality. Coaxial camera systems positioned in the reflected laser optical path enable real-time observation of melt pool geometry, including size, shape, and stability, which are critical indicators of process quality [16,17]. Building on these capabilities, advanced image processing techniques have further enhanced the capabilities of vision-based monitoring, enabling automated detection of defects such as lack of fusion, excessive penetration, and irregular bead geometry [18,19]. Nevertheless, vision-based approaches face inherent limitations including surface-only detection capabilities, sensitivity to environmental factors such as ambient light and challenges in monitoring subsurface defects that may significantly impact mechanical properties. As noted by Khanafer et al. [20], optical methods may also have limited penetration depth, particularly when inspecting opaque or highly reflective materials, which restricts their applicability for comprehensive defect detection in AM processes.

Complementing vision-based monitoring and addressing its surface-limitation constraints, Acoustic Emission (AE) monitoring has emerged as a valuable technique for capturing process dynamics that are not readily observable through visual inspection. AE sensors detect high-frequency elastic waves generated by rapid material deformation, phase transformations, and defect formation processes occurring during DED [21,22]. This sensing modality offers the distinct advantage of detecting subsurface phenomena, including crack initiation, porosity formation, and internal stress development that may not be observable visually until later stages of the build process. Furthermore, recent studies have demonstrated the effectiveness of AE monitoring in distinguishing normal and abnormal printing conditions through sophisticated signal processing and pattern recognition techniques [8,23–25]. However, AE monitoring faces challenges related to signal complexity, noise interference from the manufacturing environment, and the need for extensive signal processing to extract meaningful features.

While vision and acoustic monitoring represent the primary sensing modalities, researchers have explored various alternative sensing technologies to capture complementary aspects of the DED process. Thermal monitoring using infrared cameras and pyrometers has proven effective in tracking temperature distributions and cooling rates, providing insights into thermal history and potential residual stress development [26]. Laser scanning techniques have been employed to measure geometric accuracy and detect dimensional deviations in real-time, offering high-precision feedback for process control applications [27]. Despite these individual contributions, each modality alone provides only a partial view of the complex process dynamics, highlighting the potential benefits of integrated monitoring strategies that can capture multiple physical phenomena simultaneously.

This recognition of single-sensor limitations has led to increased research focus on multi-modal monitoring approaches. Studies have demonstrated that sensor fusion can significantly enhance defect detection capabilities by combining complementary information sources. Wu et al. [28] developed a multi-sensor fusion system integrating high-speed cameras, photodiodes, and microphones for real-time quality classification in laser powder bed fusion, achieving recognition accuracy of 97.98%, 92.63%, and 100% for high-, medium-, and low-quality samples respectively . Similarly, Petrich et al. [29] demonstrated that combining layer-wise imagery, acoustic emissions, and multi-spectral data with scan vector information achieved 98.5% accuracy in binary defect classification, with their sensitivity analysis revealing that while optical imagery contained the highest information content, additional modalities significantly improved overall classification performance. More recently, Zou et al. [30] proposed a synchronous multi-sensor monitoring approach that combines photodiode-based melt pool light intensity measurements with high-speed camera imaging, enabling real-time detection of powder melting state variations with resolution down to 30 μm thickness changes. These multi-modal approaches have consistently shown superior performance compared to single-sensor systems, providing comprehensive process understanding that individual sensors cannot achieve independently.

As monitoring systems have become more sophisticated and data-rich, the integration of machine learning techniques has become increasingly important for effective data analysis. The integration of machine learning techniques into AM quality assessment has revolutionized the ability to predict and classify defects using process monitoring data. Various ML algorithms, including support vector machines, neural networks, and ensemble methods, have been successfully applied to analyze complex sensor signals and extract meaningful patterns indicative of part quality [31,32]. These approaches have demonstrated significant capabilities in automating defect detection processes that traditionally required extensive manual inspection, with reported classification accuracies often exceeding 90% for specific defect categories. Feature extraction techniques, ranging from statistical analysis of time-domain signals to advanced frequency-domain transformations, have proven critical in translating raw sensor data into interpretable inputs for ML models, enabling effective pattern recognition across diverse manufacturing conditions [25,33].

Building upon these machine learning advances, layer-wise analysis approaches have emerged as particularly valuable methodologies for AM quality assessment [34,35], recognizing that defects often manifest and propagate through sequential layer deposition processes. Recent studies have explored the temporal evolution of process signatures across multiple layers, enabling the development of predictive models that can anticipate quality issues before they become critical [36,37]. These approaches leverage the inherent layer-by-layer nature of AM processes to build comprehensive quality profiles, allowing for early intervention and process correction.

Despite these significant advances in DED monitoring technologies, several critical limitations persist that affect comprehensive quality assessment. First, many existing ML-based monitoring systems rely on single sensing modalities, which limit their ability to capture the full complexity of AM process dynamics. Second, current approaches often focus on binary defect detection without multi-class classification capabilities. Third, many studies utilize non-contact sensing methods that may suffer from environmental noise compared to direct structural vibration measurement. Fourth, limited work exploits the layer-by-layer nature of AM for progressive quality assessment. These gaps highlight the need for a systematic framework that combines complementary sensing modalities with advanced analytical techniques, validated through controlled experimental conditions before application to real manufacturing scenarios.

## 2. Objectives and overview of this work

In response to the identified need for systematic integration of complementary sensing modalities, this research develops a multimodal sensing framework that uniquely combines contact-based AE with coaxial vision inspection for geometric feature identification and classification in DED processes. As illustrated in Figure 1, our systematic methodology progresses through five integrated levels: (1) controlled experimental design using 3mm and 5mm through-holes as validation test cases, (2) synchronized dual-sensor data acquisition combining contact-based AE monitoring with coaxial vision inspection, (3) comprehensive feature extraction from both time-frequency acoustic signatures and melt pool morphology, (4) temporal data fusion creating layer-wise quality representations, and (5) machine learning implementation comparing multiple algorithms for optimal detection and classification performance.

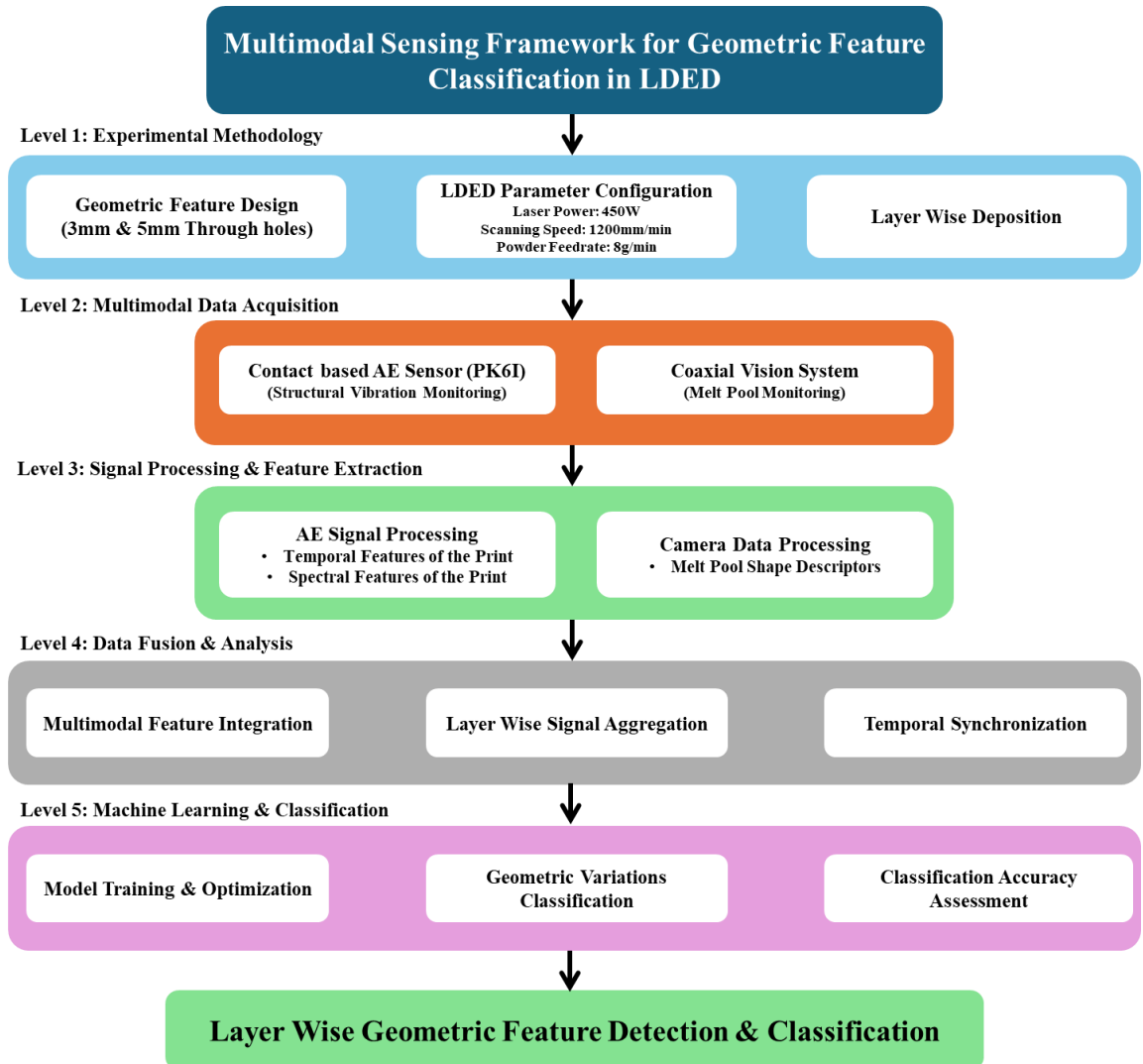

Figure 1 Five-level framework integrating experimental design, multimodal sensing, signal processing, and machine learning for geometric variation detection in DED

The primary contributions of this work include:
1. Development of a systematic framework that integrates contact-based acoustic emission and coaxial vision sensing, complete with a multi-stage pipeline for synchronized data acquisition, signal processing, and feature fusion.
2. Systematic development and comparative evaluation of multiple machine learning classifiers (e.g., SVM, Random Forest, XGBoost, and Neural Networks) to identify the optimal architecture for geometric variation classification from complex, high-dimensional sensor data.
3. Demonstration of superior, multi-class classification performance, where the optimized multimodal neural network achieved 94.4% accuracy. This result significantly surpasses single-modality approaches (87.8% for AE-only, 86.7% for Camera-only), quantitatively proving the benefits of sensor fusion.

4. Establishment of a layer-wise analysis methodology enabling early-stage detection of geometric variations from the second layer onwards. This provides the technical foundation for progressive quality assessment and future real-time intervention in AM processes.

This multimodal framework addresses the identified monitoring limitations through systematic integration of complementary sensing modalities. While this study focuses on controlled geometric variations as a validation approach, the demonstrated classification capabilities and early detection performance provide a methodological foundation for extending these techniques to manufacturing-induced defects in future work.

## 3. Experiment materials and methods

*3.1. DED system and materials*

The experimental setup employed an Optomec LENS MTS 500 (Optomec, Inc., Albuquerque, NM, USA) laser-based directed energy deposition system, which has been extensively utilized in our previous investigations of AM process monitoring [25]. This platform integrates a 500W fiber laser with precision powder delivery through four convergent nozzles, creating a focused powder stream at the deposition point. The system's closed-loop control architecture ensures consistent process parameters throughout the fabrication sequence, critical for establishing reliable correlations between sensor signatures and part quality.

Gas-atomized stainless steel 316L powder (Carpenter Additive, Philadelphia, PA, USA) with a particle size distribution of 45-106 μm served as the feedstock material. This material selection was based on its widespread adoption in industrial DED applications and well-documented thermal and mechanical properties, which facilitate reliable sensor signal generation. The powder composition adheres to ASTM A276 specifications, containing 16-18% chromium, 10-14% nickel, 2-3% molybdenum, with maximum limits of 0.03% carbon and 0.1% nitrogen. These compositional characteristics ensure predictable melting behavior and solidification dynamics throughout the deposition process. Stainless steel 316L substrates were used to ensure material compatibility and consistent acoustic wave propagation.

The controlled atmosphere was maintained through dual argon gas streams: a central shielding flow at 30 L/min protecting the melt pool from oxidation, and carrier gas at 4 L/min ensuring uniform powder delivery. The substrate-to-nozzle standoff distance was fixed at 10 mm, with converging powder streams focused to approximately 2 mm diameter at the working plane. This standoff distance provides adequate clearance for sensor integration without affecting AE signal propagation through the substrate. These parameters were established through preliminary optimization to achieve stable melt pool formation while maintaining adequate clearance for sensor integration, as detailed in the following sections.

*3.2. Experimental design and test specimens*

To systematically validate the multimodal monitoring framework, a controlled experimental design was implemented using intentionally introduced geometric features. This approach enables systematic assessment of the monitoring system's detection capabilities across varying geometric variation sizes, providing ground truth data essential for machine learning model validation. Three experimental conditions were established as shown in Figure 2:

- **Specimen without hole**: Square specimens (15mm × 15mm × 2.5mm) without geometric variations, fabricated under selected parameters to establish reference signal characteristics for both AE and vision systems.
- **3mm through-hole specimen**: Parts incorporating centrally located cylindrical through-holes with 3mm diameter. This size was selected as the minimum diameter that prevents powder accumulation within the void during deposition.
- **5mm through-hole specimens**: Parts featuring 5mm diameter through-holes, representing a larger geometric variation for comparison.

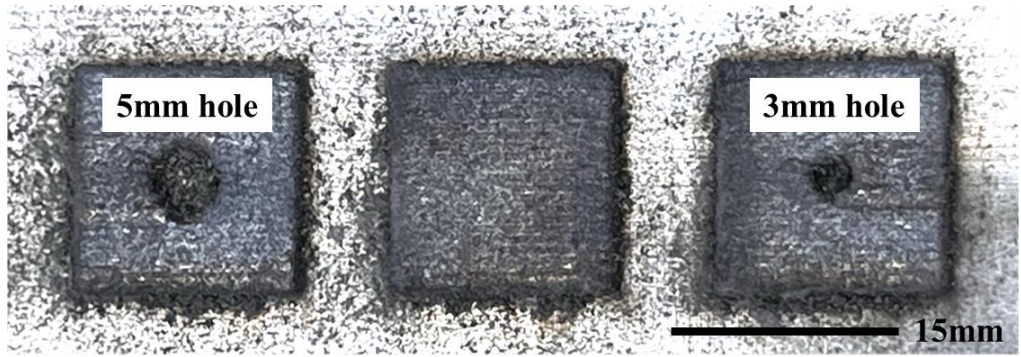

Figure 2 Three experimental conditions: 5mm through-hole specimen, Specimen without hole, 3mm through-hole specimen

Each specimen consisted of five layers deposited at 0.5mm layer height, with layers 2-5 providing independent data points for analysis. The first layer was excluded due to substrate-induced thermal boundary effects that create anomalous sensor signatures not representative of steady-state deposition. This configuration balances sufficient feature development with practical data acquisition constraints. The through-holes were integrated directly into the toolpath generation, ensuring precise geometric control.

Twenty specimens were fabricated for each experimental condition (60 total), with each specimen's layers 2-5 treated as independent observations. This yields 240 distinct data points (20 specimens × 3 conditions × 4 analyzed layers) before augmentation. The specimens were fabricated across eight build sessions with consistent positioning of both specimens and sensors. Conditions were randomized within each session to minimize systematic biases. Environmental conditions were monitored throughout (chamber temperature: $25\pm2°C$, humidity: $<40\%$) to ensure consistent sensor performance.

The layer-wise analysis approach leverages programmed 5-second pauses between successive layers. During these intervals, the deposition head retracts to a home position, creating distinct signal markers in both monitoring channels: AE amplitude drops to background levels while the vision system captures the absence of melt pool radiation. These synchronized temporal markers enable automated segmentation of continuous data streams into layer-specific datasets, as detailed in Section 4.

*3.3. Multimodal sensing system setup*

The multimodal monitoring framework integrates two complementary sensing modalities to capture both structural vibrations and surface morphology during DED. The sensor configuration, illustrated in Figure 3, was designed to maximize signal quality while maintaining practical implementation constraints.

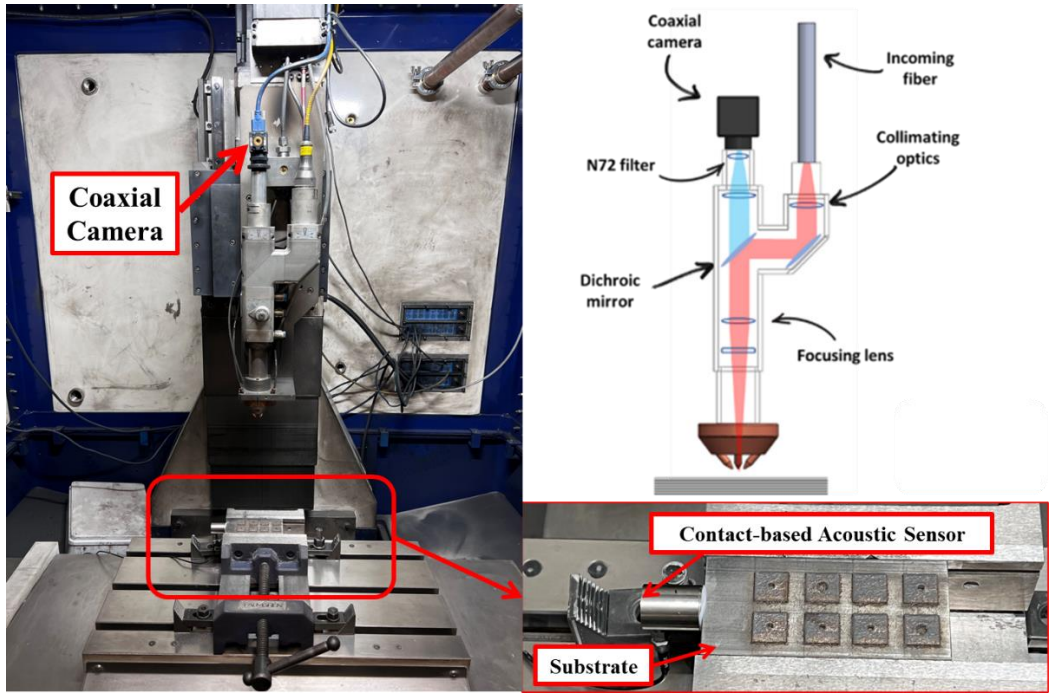

Figure 3 Multimodal sensing system configuration for DED monitoring. Left: Overall experimental setup showing the coaxial camera integration and contact-based AE sensor placement on the substrate. Right: Detailed schematic of the coaxial optical path, detailed printing layout

For AE monitoring, we employed the validated sensor configuration established in our previous work [8] where we demonstrated the effectiveness of contact based AE monitoring for DED process characterization. Specifically, a MISTRAS PK6I resonant sensor was mechanically coupled to the substrate's lateral surface using high-vacuum silicone grease, positioned 15 cm from the deposition zone as described in Section 3.2. This configuration has proven reliable for capturing high-frequency acoustic signatures associated with material deposition and defect formation [8]. As demonstrated in recent studies on optical microphones for laser process monitoring, non-contact acoustic sensors face significant challenges including limited sensitivity at higher frequencies, susceptibility to environmental noise, and signal attenuation in air [38]. In contrast, our contact-based approach provides direct mechanical coupling to the substrate, enabling high-fidelity capture of structure-borne acoustic emissions generated during the deposition process. The integrated 26 dB preamplifier maintains a low noise level (<3 $\mu V$ RMS), further enhancing the signal-to-noise ratio.

The optical monitoring subsystem utilizes a coaxial vision configuration that has been successfully implemented in our laboratory's previous DED monitoring study [14]. Building on these established methods, we employed a FLIR Blackfly S camera equipped with a Sony IMX273 CMOS sensor. The optical path incorporates an R72 near-infrared filter (>720 nm transmission) to suppress laser reflection while transmitting melt pool thermal radiation. A dichroic mirror positioned at 45° directs the incident laser beam while allowing backward-propagating thermal emissions to reach the camera sensor. This coaxial arrangement eliminates perspective distortions inherent in off-axis configurations and provides a consistent 3.6 × 3.6 mm field of view at 4.5 µm/pixel resolution, sufficient for detecting morphological variations induced by geometric features.

Both sensing systems operate independently during data acquisition, with the AE system sampling continuously at 500 kHz using a 32 dB threshold and the vision system acquiring images at 30 fps. Temporal alignment between modalities is achieved during post-processing through identification of common layer boundaries. The programmed 5-second dwell periods create distinctive markers in both data streams—AE amplitude drops below 30 dB while melt pool radiation ceases—enabling precise layer-wise segmentation and alignment. This methodology enables direct correlation between acoustic events and visual phenomena, providing the foundation for the multimodal analysis framework presented in Section 4.

### 3.4. Experimental Protocol and Data Acquisition

The experimental procedures followed a systematic protocol building upon the established sensor configuration and specimen design. Prior to specimen fabrication, comprehensive calibration procedures were conducted to establish optimal sensor operating conditions. The camera system underwent geometric calibration using a checkerboard pattern to correct lens distortion, followed by exposure optimization to prevent melt pool image saturation while maintaining adequate sensitivity. For the AE system, signal responses were characterized under four distinct operational modes: mechanical movement only, powder flow without laser, laser operation without powder, and complete printing conditions. This systematic characterization enabled identification of process-relevant acoustic signatures and establishment of the 32 dB acquisition threshold that effectively discriminates between manufacturing signals and environmental noise.

All specimens were subsequently fabricated using consistent DED parameters: laser power of 450 W, scanning speed of 1200 mm/min, and powder feed rate of 6 RPM. These parameters remained constant throughout the entire study to isolate the effects of geometric variations on sensor signatures. The layer-wise fabrication sequence incorporated the critical 5-second dwell periods between layers, during which the laser deactivated, and the deposition head retracted. These programmed pauses generated the temporal markers essential for automated data segmentation in subsequent analysis.

Continuous data acquisition proceeded throughout each print session, with each sensing system recording independently. The controlled introduction of geometric variations (3mm and 5mm through-holes) through toolpath programming ensured that all specimens maintained their designed dimensions as verified by the CAD-based toolpath generation as shown in Figure 2. This experimental framework, progressing from calibrated sensor setup through controlled fabrication to comprehensive data collection, establishes the foundation for the multimodal analysis presented in Section 4. The layer-wise data structure, combined with precise geometric ground truth from the programmed variations, enables quantitative assessment of the monitoring system's detection and classification capabilities.

## 4. Data processing and feature extraction

*4.1.Acoustic emission signal processing*

The AE data processing transforms the continuous waveforms captured by the contact-based sensor system into meaningful features that characterize the DED process state. Building upon the synchronized data acquisition framework established in Section 3.3, the raw AE signals undergo systematic processing to extract meaningful descriptors.

Figure 4 illustrates the complete AE signal spanning an entire fabrication sequence. The prominent amplitude reductions correspond to the 5-second dwell periods programmed between layers, as described in Section 3.2. These quiet zones, where signal amplitude drops, serve as temporal markers for layer change detection. The segmentation algorithm employs a dual-threshold approach: when the signal amplitude remains below 30 dB for more than 4.5 seconds, a layer transition is registered. This threshold was empirically determined through analysis of multiple build sequences to reliably distinguish dwell periods from transient signal variations.

Following segmentation, each layer's data undergoes temporal trimming to ensure consistent analysis windows across all specimens. The initial 10 seconds and final 5 seconds of each layer are systematically removed, as shown in Figure 4(a). This standardized trimming window was selected to exclude potential edge effects while ensuring the central region containing geometric variations remains fully captured. This preprocessing step reduces the dataset size by approximately 30%, facilitating computational efficiency.

The trimmed signals then undergo noise reduction through outlier detection and management. Figure 4(b) demonstrates the application of Median Absolute Deviation (MAD) analysis to identify statistical outliers, signal points exceeding $\pm 3\sigma$ from the local median. Winsorization is applied to cap these extreme values at the $\pm 3\sigma$ boundaries rather than eliminating them entirely, preserving signal dynamics while limiting the influence of noise spikes on subsequent calculations. High-pass filtering constitutes the next preprocessing step, as illustrated in Figure 4(c). A Butterworth filter with a 150 kHz cutoff frequency is applied to isolate high-frequency components. This cutoff frequency was selected based on preliminary FFT analysis comparing the three experimental conditions, which revealed that spectral differences between geometric variations were most pronounced above 150 kHz. The filtered signal (shown in green) retains rapid transients while eliminating low-frequency baseline variations evident in the preprocessed signal.

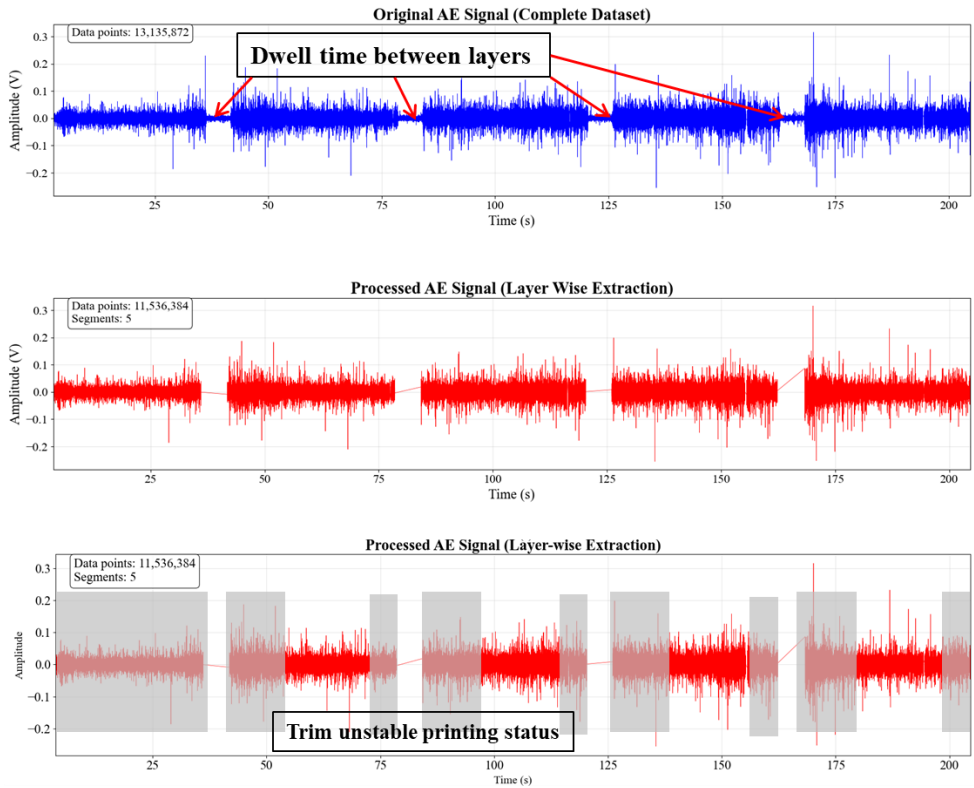

(a) Layer boundary detection and stable region extraction from continuous AE signals

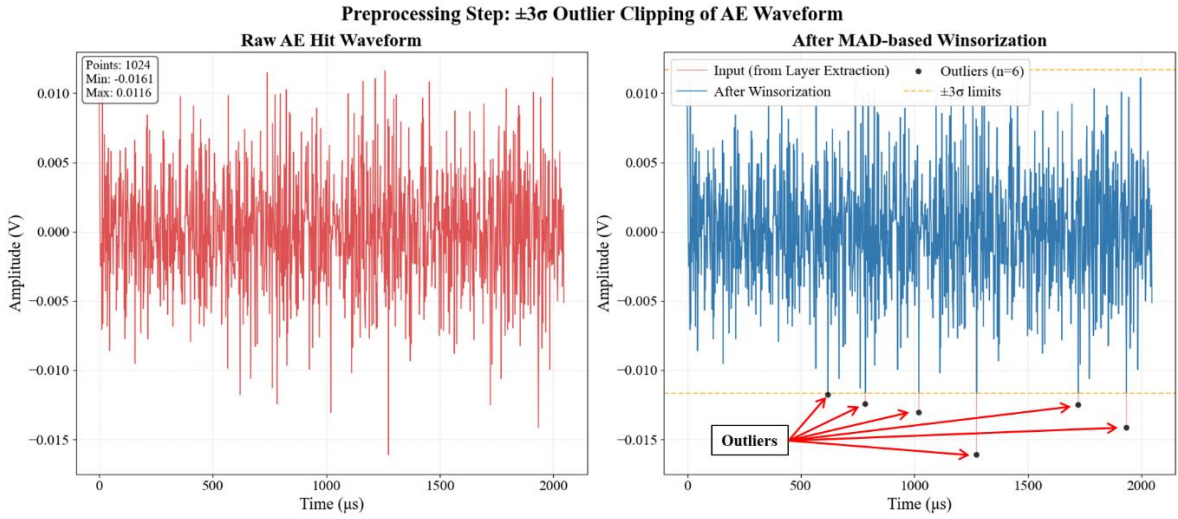

(b) Outlier detection and Winsorization preprocessing of AE waveforms

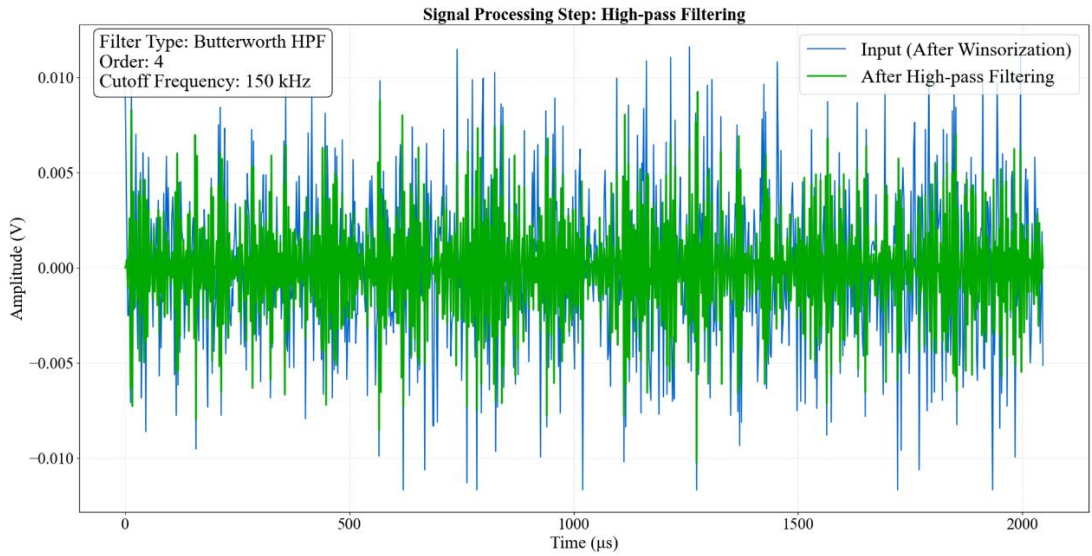

(c) High-pass filtering for isolation of high-frequency AE components

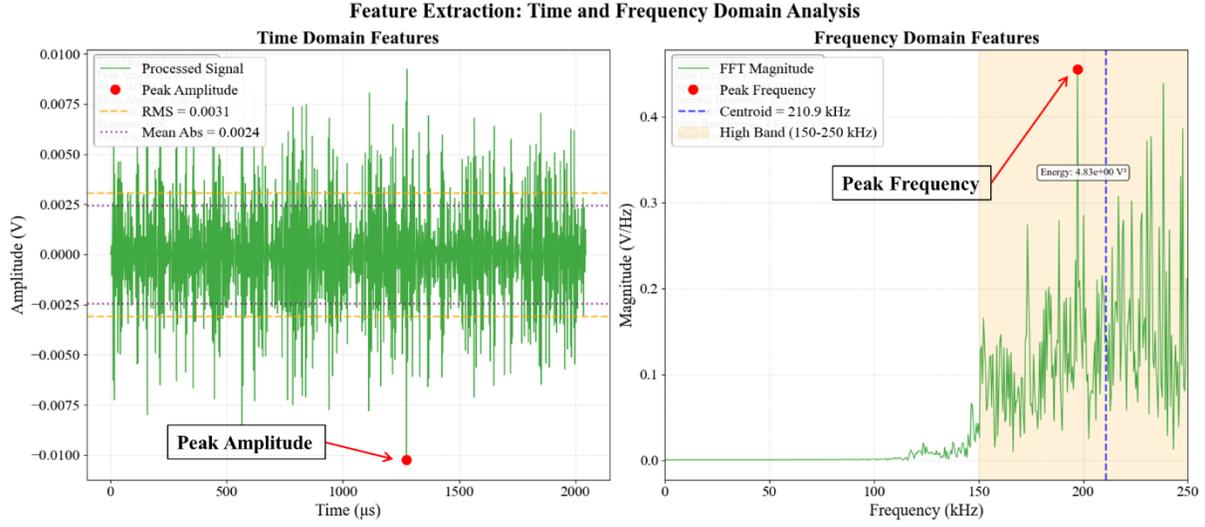

(d) Time-domain and frequency-domain feature extraction from processed AE signals

Figure 4 AE signal processing and feature extraction pipeline (a) Layer boundary detection and stable region extraction from continuous AE signals (b) Outlier detection and Winsorization preprocessing of AE waveforms (c) High-pass filtering for isolation of high-frequency AE components (d) Time-domain and frequency-domain feature extraction from processed AE signals

For feature extraction, the preprocessed signals are segmented into analysis windows of 1024 sampling points (2.048 ms at 500 kHz sampling rate) as illustrated in Figure 4(d). The comprehensive feature extraction process generates 72 distinct characteristics—36 from filtered signals and 36 from raw signals—for each analysis window. These features span both time-domain and frequency-domain analyses to capture the complete acoustic signature of geometric variations.

Statistical features form the core of the time-domain analysis. Kurtosis, which measures the "peakedness" of the signal amplitude distribution, is calculated as:

$$\text{Kurtosis} = \frac{E[(X-\mu)^4]}{\sigma^4} \qquad (1)$$

where N=1024 samples per window, $x_i$ represents the signal amplitude at sample $i$, $\mu$ is the mean amplitude, and $\sigma$ is the standard deviation. This metric proves particularly sensitive to transient events in the acoustic emissions.

Mean absolute amplitude (MAA) quantifies the average signal strength:

$$\text{Mean Absolute Amplitude (MAA)} = \frac{1}{N}\sum_{i=1}^{N}|x_i| \qquad (2)$$

while its standard deviation captures amplitude variability within each window.

$$\text{Standard Deviation} = \sqrt{\frac{1}{N-1} \sum_{i=1}^{N} (x_i - \mu)^2} \tag{3}$$

Energy-based features provide complementary information about signal power. The RMS (root mean square) value represents the signal's overall power content:

$$\text{RMS} = \sqrt{\frac{1}{N} \sum_{i=1}^{N} x_i^2} \tag{4}$$

Absolute energy is computed as the sum of squared amplitudes:

$$E_{abs} = \sum_{i=1}^{N} x_i^2 \tag{5}$$

Band energy, calculated for specific frequency ranges, quantifies spectral power distribution:

$$E_{band} = \sum_{f=f_1}^{f_2} |X(f)|^2 \tag{6}$$

where $X(f)$ represents the Fourier transform of the signal and $f_1, f_2$ define the frequency band boundaries.

Frequency-domain features are extracted following Fast Fourier Transform of each window. Key spectral features include the energy ratio between low and high frequency bands:

$$ER_{lowhigh} = \frac{\sum_{f=0}^{f_{mid}} |X(f)|^2}{\sum_{f=f_{mid}}^{f_{max}} |X(f)|^2} \tag{7}$$

and the spectral centroid, which indicates the "center of mass" of the spectrum:

$$f_{centroid} = \frac{\sum_f f \cdot |X(f)|^2}{\sum_f |X(f)|^2} \tag{8}$$

Additional features include band energy for specific frequency ranges, spectral entropy, bandwidth, and peak frequencies. This comprehensive feature set, totaling 72 metrics per window (36 each from filtered and raw signals), captures both temporal and spectral characteristics of the acoustic emissions. Statistical aggregation of these window-based features into layer-wise representations ensures that both high-frequency transients and broader process dynamics are preserved in the final feature vectors.

This multi-stage signal processing pipeline—progressing from raw waveforms through segmentation, filtering, and comprehensive feature extraction—transforms the continuous AE data stream into structured descriptors suitable for machine learning analysis. While these acoustic features provide valuable insights into the structural dynamics and subsurface phenomena occurring during material deposition, they capture only one aspect of the complex DED process. Surface-level information, particularly the evolution of melt pool morphology that directly reflects the interaction between laser energy and material, remains equally critical for comprehensive geometric variation characterization. The following section details the parallel processing pipeline developed for extracting these complementary visual features from the coaxial camera system.

*4.2. Melt pool analysis*

The coaxial camera system captures the melt pool irradiance throughout the print process, providing complementary surface-level information to the subsurface phenomena detected through AE. The camera data processing parallels the systematic approach established for AE analysis, transforming raw thermal images into quantitative geometric descriptors.

Layer boundary detection in the camera data stream leverages the cessation of melt pool radiation during dwell periods. As illustrated in Figure 5(a), the raw melt pool intensity exhibits sharp drops to near-zero values when the laser turns off between layers. The algorithm identifies layer transitions when consecutive frames show pixel intensities below 10% of the mean active printing intensity for more than 4.5 seconds, consistent with the temporal markers used in AE processing. The detected layer boundaries (shown as red dots) enable automated segmentation of the continuous image stream into layer-specific datasets.

Following layer identification, the vision data undergoes temporal trimming analogous to the AE processing pipeline. The lower panel of Figure 5(a) demonstrates the extracted layer-wise data after removing the initial 10 seconds and final 5 seconds of each deposition sequence, as indicated by the gray shaded regions. This preprocessing ensures consistency with the AE data structure while focusing analysis on stable melt pool conditions.

The preprocessed image frames then undergo morphological analysis to extract quantitative melt pool characteristics. Figure 5(b) illustrates the multi-stage processing pipeline applied to each frame. Beginning with the raw thermal image captured through the R72 near-infrared filter, intensity-based thresholding segments the high-temperature region from the background. The threshold value is adaptively set at 80% of the maximum frame intensity to accommodate variations in overall image brightness across different build sessions. Following segmentation, morphological operations refine the melt pool boundary and eliminate spatter-induced artifacts using the validated approach detailed in [14].

From the segmented melt pool, geometric features are extracted as demonstrated in Figure 5(b). The minimum enclosing circle algorithm provides fundamental size metrics including circle area and radius.

Shape descriptors quantify morphological characteristics commonly used in melt pool analysis. The core-to-circle ratio measures how well the melt pool fills its enclosing circle:

$$\text{Core 2 Circle} = \frac{A_{\text{contour}}}{A_{\text{circle}}} \tag{9}$$

where $A_{contour}$ represents the actual melt pool area. Values approaching 1 indicate circular melt pools, while lower values suggest irregular or elongated shapes. Convexity measures boundary regularity:

$$\text{Convexity} = \frac{A_{\text{contour}}}{A_{\text{convex}}} \tag{10}$$

where $A_{convex}$ is the area of the convex hull enclosing the melt pool. The bounding box dimensions (Length and Width) characterize the overall melt pool extent in the principal directions.

To create robust layer-wise representations, statistical aggregation is applied—computing both mean and standard deviation for each geometric feature across all frames within a layer. This approach captures not only typical melt pool characteristics but also their stability throughout deposition. The complete feature set comprises standard geometric metrics including contour area, circle properties, convexity, and bounding box dimensions, along with layer-wise frame count and time span information.

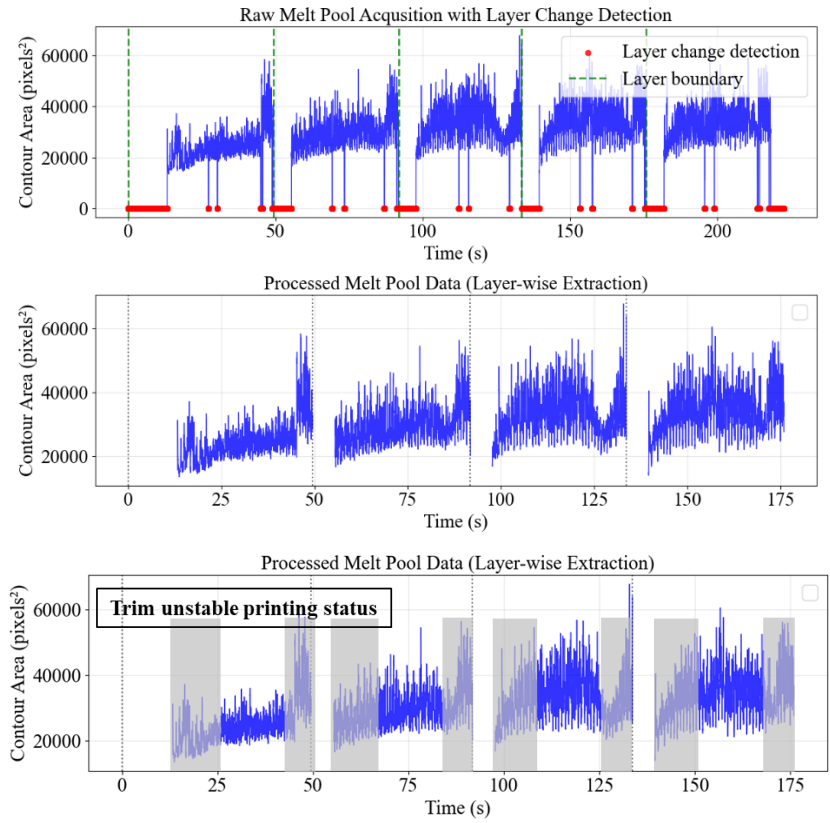

(a): Raw melt pool acquisition showing layer change detection markers and processed layer-wise extraction with temporal trimming of unstable printing states

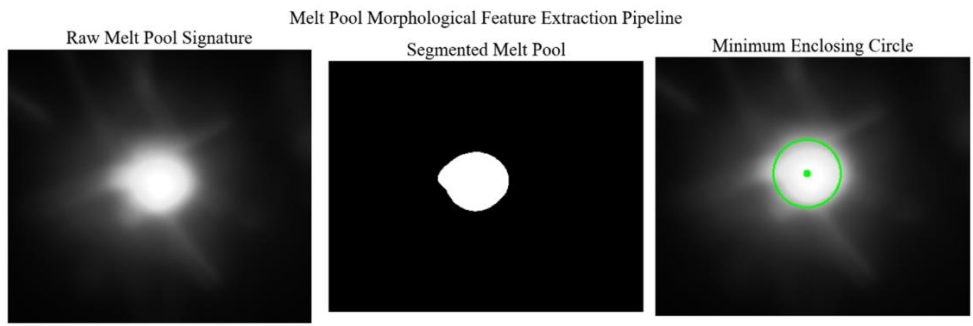

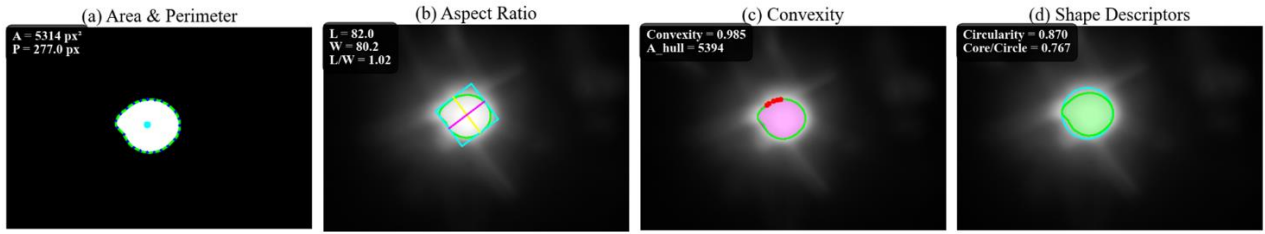

(b)

Figure 5 Vision-based melt pool analysis pipeline. (a) Raw melt pool acquisition showing layer change detection markers and processed layer-wise extraction with temporal trimming of unstable printing states. (b) Melt pool morphological feature extraction pipeline showing raw thermal signature, binary segmentation, and geometric characterization including area/perimeter measurements, aspect ratio, convexity, and shape descriptors.

These derived features provide essential complementary information to the AE analysis. The integration of both sensing modalities through the multimodal framework in Section 4.3 enables comprehensive geometric variation characterization—combining subsurface structural information from AE with surface-level geometric variations from vision monitoring.

### 4.3. Data fusion Augmentation and multimodal feature engineering

The individual processing pipelines detailed in Sections 4.1 and 4.2 yield comprehensive feature sets from AE and camera monitoring. While each modality provides valuable insights—AE capturing subsurface structural dynamics and vision monitoring surface morphology—their true diagnostic potential emerges through systematic integration. This section presents the hierarchical framework that transforms these parallel data streams into a unified multimodal representation.

Figure 6 illustrates the five-level integration framework. At the foundation (Level 1), time-domain and frequency-domain features from AE signals are combined with geometric features from vision analysis, along with ground truth labels corresponding to the three experimental conditions (normal, 3mm hole, 5mm hole) established in Section 3.2. The AE features include both filtered and raw signal processing to capture high-frequency transients and full-spectrum dynamics, while vision features encompass melt pool geometric descriptors detailed in Section 4.2.

Building upon the layer-wise data structure introduced in Section 3.2, Level 2 achieves temporal synchronization through the programmed dwell periods. These 5-second pauses create distinctive markers in both modalities—AE amplitude drops below the 30 dB threshold while simultaneously melt pool radiation ceases in vision data. This dual-sensor validation of layer boundaries enables precise signal alignment and specimen-layer mapping, eliminating the need for complex cross-correlation techniques.

With aligned multimodal data established, Level 3 implements feature selection through statistical analysis. The ANOVA F-test evaluates each feature's ability to discriminate between geometric variation categories, creating a unified ranking across both AE and vision features. This dimensionality reduction retains the most informative features from both modalities while creating a computationally tractable multimodal feature set, with detailed selection results presented in Section 6.1.

Level 4 transforms the variable-length sequences into fixed-dimensional layer-wise representations. Statistical aggregation computes mean and standard deviation for each feature, whether from AE windows or vision frames within a layer. Z-score standardization subsequently normalizes the combined feature matrix, ensuring balanced contributions from both sensing modalities despite their different physical units and magnitudes.

The final level implements data augmentation to enhance model training. SMOTE interpolation generates synthetic samples by combining features from similar specimens, Gaussian perturbation adds controlled noise simulating measurement variability in both sensors, and class balancing ensures equal representation across all geometric variation categories. Critically, these augmentation strategies preserve the layer-wise structure and physical relationships between acoustic and visual features.

Through this systematic five-level framework, the raw AE waveforms and thermal images are transformed into an integrated multimodal dataset optimized for machine learning analysis. The following section details how this unified representation enables robust classification of geometric variations across multiple algorithmic approaches.

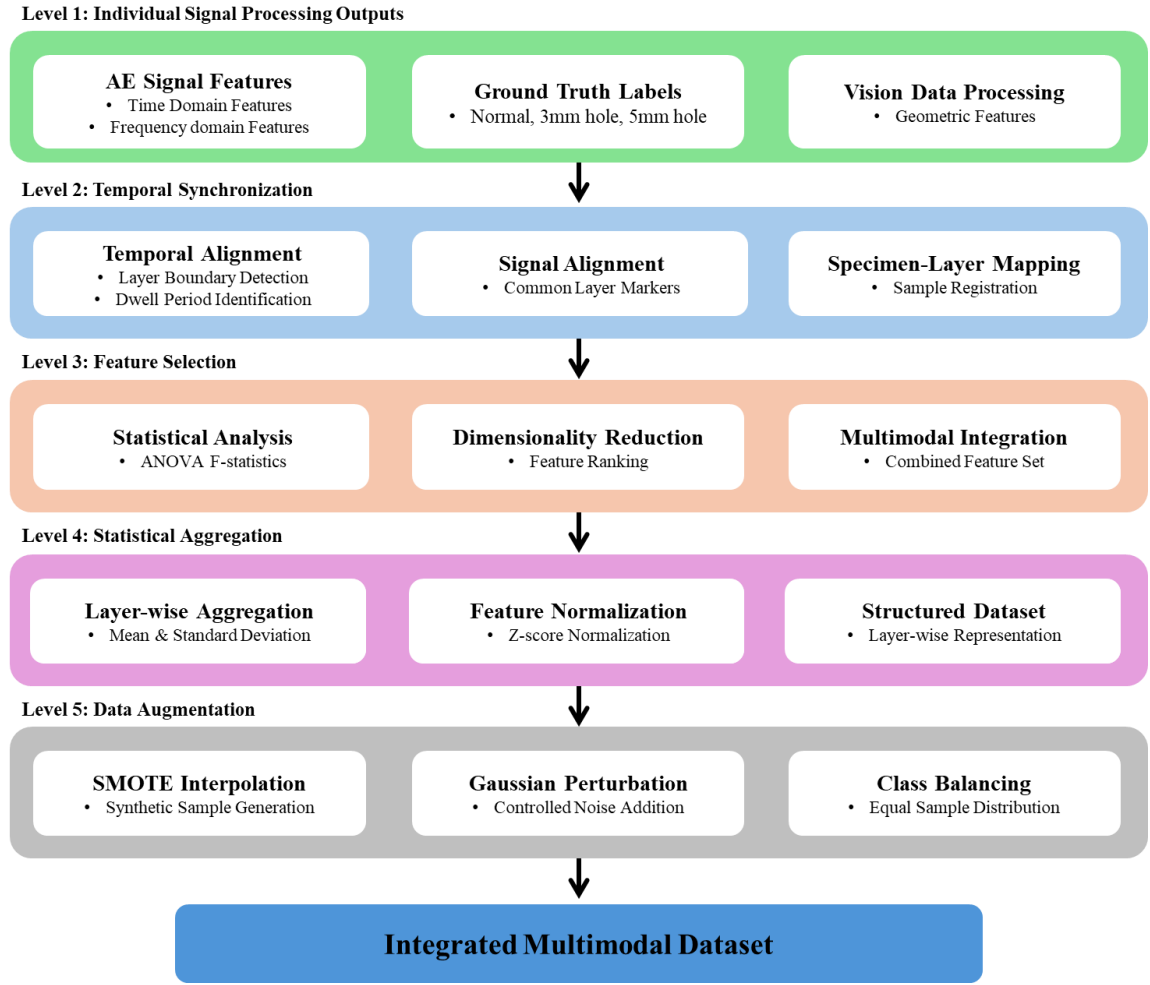

Figure 6 Multimodal data integration and preprocessing framework

## 5. Machine Learning Models Development

*5.1. Model architecture selection and comparison*

The multimodal dataset prepared through the systematic integration framework (Section 4.3) provides the foundation for developing machine learning models capable of detecting and classifying geometric variations. With comprehensive features capturing both acoustic and visual phenomena, the modeling objective focuses on identifying architectures that can effectively leverage multimodal data while maintaining generalization capability despite the limited experimental dataset size.

The model selection strategy evaluated six distinct classifier architectures, chosen to represent different learning approaches and complexity levels. This comprehensive evaluation ensures robust performance assessment across various algorithmic approaches, with each architecture tested on both individual modalities and multimodal fusion to systematically quantify the benefits of sensor integration.

Neural networks were selected as the primary deep learning approach due to their ability to learn complex non-linear relationships between features. Flexible architecture allows separate processing pathways for acoustic and visual features before fusion, making them particularly suitable for the multimodal data structure established in Section 4.3. Through multiple hidden layers, these models can discover hierarchical patterns that may not be evident in the original feature space.

Support Vector Machines (SVM) with radial basis function kernels provide an alternative approach by mapping features into high-dimensional spaces where geometric variations become linearly separable. This transformation enables effective classification even with limited training samples, addressing a key constraint in DED monitoring where each experimental specimen requires significant fabrication time and resources.

Tree-based ensemble methods offer complementary classification strategies. Random Forest aggregates predictions from multiple decision trees trained on different data subsets, inherently reducing overfitting through averaging. Gradient Boosting takes a sequential approach, with each tree correcting residual errors from its predecessors. XGBoost extends this framework with additional regularization to prevent overfitting, particularly important given our augmented dataset structure. These methods naturally capture feature interactions through recursive partitioning, potentially revealing synergies between acoustic signatures and visual characteristics.

Logistic Regression serves as the linear baseline, testing whether the engineered features from Section 4.3 create linearly separable patterns. Despite its simplicity, L2-regularized logistic regression can achieve competitive performance when feature engineering effectively captures the underlying physics—as demonstrated by the discriminative features identified through ANOVA analysis.

This algorithmic selection enables systematic evaluation across multiple dimensions. Linear versus non-linear models test the complexity of decision boundaries required for geometric variation classification. Individual classifiers versus ensemble methods evaluate whether prediction aggregation improves robustness. Simple architecture versus deep networks assesses the trade-off between model complexity and generalization capability on our limited experimental dataset.

Beyond classification performance, tree-based methods provide interpretability through feature importance rankings, revealing which acoustic or visual measurements drive predictions. This comprehensive evaluation framework, applied to both single-modality and multimodal configurations, quantifies the benefits of sensor fusion while identifying the most effective learning approach for real-time DED monitoring applications. The following sections detail the hyperparameter optimization process and resulting classification performance.

*5.2. Hyperparameter optimization and training strategy*

Following the selection of six classifier architectures, systematic hyperparameter optimization was conducted to identify optimal configurations for each model. The optimization process explored predefined parameter ranges to balance model performance with computational efficiency.

For neural networks, architectural parameters were optimized including the number of hidden layers (2-4 layers), neurons per layer (16-128), dropout rates (0.15-0.35), and L2 regularization strength (0.001-0.01). Learning rates were searched within the range of 0.0001-0.001, with early stopping patience varying from 10-30 epochs based on validation performance.

Traditional machine learning models underwent grid search optimization across the following parameter ranges:

- SVM: Regularization parameter C (0.1-10) and RBF kernel gamma ('scale', 'auto', 0.001-0.1)
- Random Forest: Number of estimators (30-200), maximum depth (3-15), minimum samples split (2-10)
- Gradient Boosting: Number of estimators (30-200), learning rate (0.01-0.2), maximum depth (3-8), subsample ratio (0.6-1.0)
- XGBoost: Number of estimators (30-150), learning rate (0.01-0.1), maximum depth (2-6), subsample and column sampling ratios (0.5-0.8)
- Logistic Regression: Regularization parameter C (0.01-1.0) with L2 penalty

The optimization process identified configurations that achieved optimal performance across all modalities. The selected parameters were then applied consistently to both single-modality and multimodal configurations to ensure fair comparison. All models were trained on standardized features using z-score normalization, with the `StandardScaler` object fitted exclusively on training data. This systematic approach ensures that performance differences observed in the evaluation reflect the inherent value of sensor fusion rather than parameter selection bias.

*5.3. Model evaluation and performance assessment*

The optimized models underwent comprehensive evaluation to assess their effectiveness in detecting and classifying geometric variations. The evaluation framework examined multiple performance dimensions relevant to real-time DED monitoring applications.

Performance evaluation employed stratified train-test splitting (85/15) to preserve the class distribution established through the augmentation process described in Section 4.3. This split maintained balanced representation across all three geometric variation categories, ensuring unbiased evaluation across experimental conditions.

Classification performance was quantified through four complementary metrics. Accuracy measured overall correct predictions across all geometric categories. Precision assessed the reliability of variation detection—critical for minimizing false alarms in production settings. Recall evaluated sensitivity to actual geometric variations, ensuring minimal missed detections. The F1-score provided a harmonic mean of precision and recall, offering a balanced assessment particularly relevant for quality control applications.

To systematically evaluate the contribution of each sensing modality, three input configurations were tested. Section 6.1 details the ANOVA-based feature selection that identified the most discriminative features for each modality. The configurations included AE-only using acoustic features, camera-only using vision-based geometric features, and multimodal fusion combining both feature sets. All six classifier architectures were trained and evaluated on each configuration using identical data splits and preprocessing pipelines. This controlled ablation study enabled direct quantification of sensor fusion benefits.

The evaluation framework also incorporated temporal analysis to assess detection capability evolution throughout the build process. By examining performance across layers 2-5 independently, the analysis revealed how geometric variation signatures develop during deposition. This layer-wise assessment addresses the critical requirement for early detection in real-time process control applications.

For neural networks, model initialization effects were evaluated through multiple training runs, with performance statistics computed across iterations. The evaluation protocol ensured that reported metrics represent robust model behavior under consistent experimental conditions.

This comprehensive evaluation framework provides systematic assessment across three key dimensions: comparative analysis of classifier architectures, quantification of multimodal fusion benefits through ablation studies, and temporal characterization of detection capabilities. The detailed results presented in Section 6 demonstrate the practical viability of the proposed monitoring approach for geometric variation classification in DED manufacturing.

## 6. Result and discussion

*6.1. Sensor feature analysis and characterization*

*6.1.1. Acoustic emission feature analysis*

The comprehensive feature extraction framework established in Section 4.1 generated 72 AE features from each analysis window—36 from filtered signals (>150 kHz) and 36 from raw signals. This dual processing strategy enabled capture of both high-frequency geometric variation signatures and full-spectrum process dynamics, providing a rich dataset for identifying the most discriminative characteristics.

Feature selection using Analysis of Variance (ANOVA) F-test quantified each feature's ability to discriminate between the three experimental conditions. The F-statistic measures the ratio of between-group to within-group variance:

$$F = \frac{MSB}{MSW} = \frac{\sum_{i=1}^{k} n_i (\bar{x}_i - \bar{x})^2 / (k-1)}{\sum_{i=1}^{k} \sum_{j=1}^{n_i} (x_{ij} - \bar{x}_i)^2 / (N-k)} \tag{11}$$

where $k$ represents the experimental conditions (part without hole, part with 3mm hole, part with 5mm hole), $n_i$ is the number of samples in group $i$, $N$ is the total sample size, and larger F-values indicate stronger discriminative capability with correspondingly smaller p-values.

Table 1 presents the five most discriminative acoustic features ranked by F-statistic. Kurtosis-based features demonstrate exceptional discriminative power, with filtered kurtosis mean achieving the highest F-value (317.3, $p < 1.36 \times 10^{-94}$). As defined in Section 4.1, kurtosis quantifies the "peakedness" of signal amplitude distributions, with higher values indicating peaked distributions with heavy tails. This characteristic reflects the transient acoustic events generated when the deposition process encounters geometric discontinuities, aligning with the physical understanding that geometric variations induce sudden changes in structural vibration patterns.

Energy-based features constitute the remaining top-ranked metrics, with mean absolute amplitude standard deviation (F = 165.5) and absolute energy standard deviation (F = 165.3) showing nearly identical discriminative capabilities. These features capture amplitude variability during deposition, suggesting that geometric variations not only alter signal intensity but also introduce inconsistencies in the AE patterns. Notably, all top five features derive from either filtered signals or standard deviation measurements, confirming that high-frequency components and signal variability provide the most reliable indicators of geometric variations.

Table 1 Top-ranked AE features with F-statistic and p-values

| Rank | Feature | F-statistic | p-value |
| --- | --- | --- | --- |
| 1 | Mean Kurtosis | 317.257 | 1.36E-94 |
| 2 | STD of Mean Absolute Amplitude | 165.472 | 6.65E-58 |
| 3 | STD of Absolute Energy | 165.297 | 7.45E-58 |
| 4 | STD of Band Energy | 158.249 | 7.20E-56 |
| 5 | STD of RMS | 151.542 | 5.96E-54 |

Figure 7 presents the distribution analysis of the top five discriminative features across experimental conditions. In these box plots, the central line indicates the median, the box boundaries represent the first and third quartiles (Q1 and Q3), the whiskers extend to values within 1.5 times the interquartile range. These features can be categorized into two distinct groups based on their response patterns.

The filtered mean value of kurtosis shows normal specimens with the lowest values (median ≈ 2.98), 5mm holes with intermediate values (median ≈ 3.00), and 3mm holes with the highest values (median ≈ 3.01). This non-monotonic relationship indicates that kurtosis response varies with hole size in a non-linear manner. The remaining four features, all standard deviation metrics of energy-based parameters, demonstrate a different pattern. These variability measures show the lowest values for 3mm holes, intermediate values for normal specimens, and highest values for 5mm holes. For instance, absolute energy standard deviation measures approximately 0.94 V·s for 3mm holes, 1.08 V·s for normal specimens, and 1.15 V·s for 5 mm holes.

The contrasting patterns between kurtosis (highest for 3mm holes) and energy variability metrics (lowest for 3mm holes) demonstrate that different AE characteristics respond differently to geometric variations. While kurtosis peaks at the intermediate void size, energy-based variability measures show a V-shaped response with minimum at 3mm. These distinct response patterns across different feature types justify the comprehensive feature extraction approach and contribute to the high discriminative power observed in the ANOVA analysis.

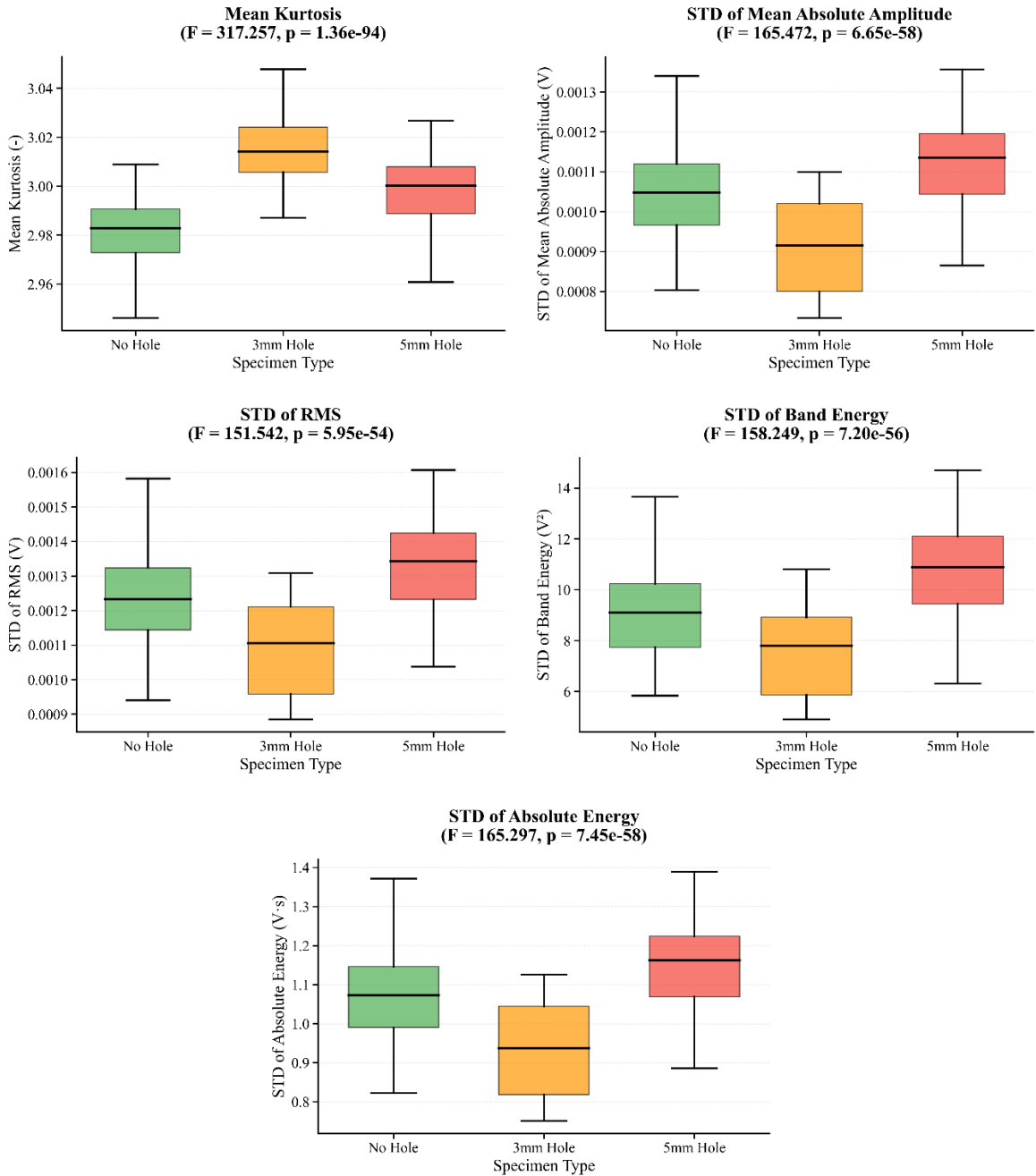

Figure 7 Box plots of discriminative AE features for different geometric conditions. Four plots show standard deviation (STD) values of different AE parameters and one shows mean kurtosis, comparing No Hole, 3mm Hole, and 5mm Hole specimen types.

Beyond the distributions examined in Figure 7, understanding how these acoustic signatures evolve during the build process is essential for real-time monitoring applications. Figure 8 presents the layer-wise progression of representative features across layers 2-5, revealing how geometric variation signatures develop and persist throughout deposition.

Mean kurtosis evolution maintains its distinct behavior as a shape-based metric. All three conditions show clear separation throughout the print, with 3mm holes consistently exhibiting the highest values (3.009 to 3.022), normal specimens the lowest (2.984 to 2.966), and 5mm holes intermediate values. Unlike the energy-based features, kurtosis demonstrates remarkable stability across layers—the rank ordering established in layer 2 persists through layer 5. This temporal consistency reinforces kurtosis as a reliable early indicator of geometric variations, detectable from the initial analysis layer.

The four energy variability metrics exhibit similar evolutionary patterns that differ markedly from kurtosis behavior. While 3mm holes start with the lowest values at layer 2 across all four metrics, they show dramatic increases to layer 3, surpassing normal specimens. For instance, absolute energy standard deviation for 3mm holes jumps from 0.80 V·s (layer 2) to 1.06 V·s (layer 3), exceeding the normal specimen value of 1.01 V·s. The subsequent evolution varies by condition. For normal specimens and 5mm holes, most features peak at layer 4—RMS standard deviation reaches maximum values of 0.00139 V (normal) and 0.00142 V (5mm) at this layer. In contrast, 3mm holes typically peak earlier at layer 3, then decline through layers 4 and 5. This results in a characteristic pattern where 3mm holes transition from lowest values (layer 2) to intermediate values (layer 3), before returning to lowest values by layer 5.

The contrasting temporal patterns between kurtosis (maintaining consistent separation) and energy variability metrics (showing dynamic crossovers and peaks) demonstrate that these feature categories respond differently to the same geometric variations. This divergent behavior justifies the multi-feature extraction approach, as no single feature type fully characterizes the acoustic response across all layers. The ability to detect geometric variation signatures from layer 2 onwards, as evidenced by the clear separations in both feature categories, supports the feasibility of early-stage quality monitoring. These layer-wise AE characteristics provide the temporal feature evolution data essential for training the machine learning models discussed in subsequent sections.

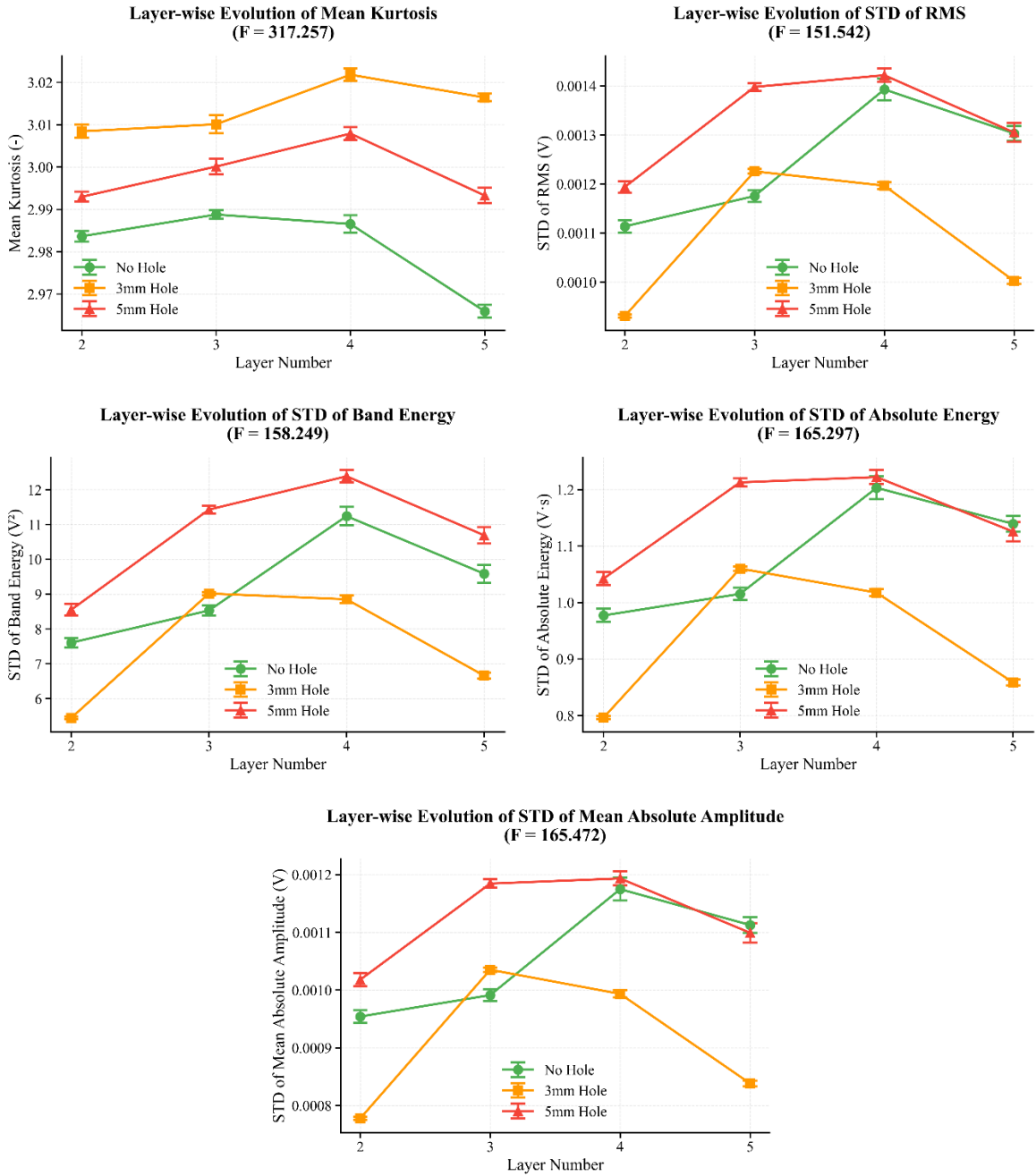

Figure 8 Layer-wise evolution of discriminative AE features for different geometric conditions

### 6.1.2. Coaxial camera system feature analysis

Complementing the AE analysis, the vision-based monitoring system captured melt pool characteristics throughout deposition using the processing framework established in Section 4.2. The 15 extracted features underwent the same ANOVA F-test analysis to identify those most sensitive to geometric discontinuities.

Table 2 presents the top-ranked vision features, revealing distinct patterns compared to the acoustic results. The mean value of core-to-circle ratio emerges as the most discriminative feature ($F = 82.5, p < 2.34 \times 10^{-32}$), quantifying deviations from circular melt pool symmetry. As defined in Section 4.2, this metric represents how well the actual melt pool area fills its minimum enclosing circle, with values approaching 1 indicating more circular shapes.

Table 2 Top-ranked vision features with F-statistic and p-values.

| Rank | Feature | F-statistic | p-value |
|---|---|---|---|
| 1 | core2circle_ratio_mean | 82.487 | 2.34E-32 |
| 2 | circle_area_std | 29.148 | 8.35E-13 |
| 3 | convexity_mean | 29.017 | 9.41E-13 |
| 4 | circle_radius_std | 29.011 | 9.46E-13 |

Figure 9 presents the distribution analysis of representative vision features across experimental conditions. These features can be categorized into two groups based on their response patterns: shape regularity metrics and size variability metrics.

The shape regularity metrics—core-to-circle ratio mean and convexity mean—both show highest values for 3mm holes, indicating more regular and circular melt pools for this intermediate void size. Core-to-circle ratio increases from 0.795 (normal) to 0.815 (3mm holes), while convexity shows a similar pattern. The 5mm holes exhibit intermediate values between normal specimens and 3mm holes for both metrics.

In contrast, the size variability metrics—circle radius standard deviation and circle area standard deviation—demonstrate an inverse pattern. Normal specimens show the highest variability in melt pool size (radius STD ≈ 6.5 mm, area STD ≈ 1900 mm²), while 3mm holes exhibit the lowest variability (radius STD ≈ 5.0 mm, area STD ≈ 1300 mm²). The 5mm holes show intermediate variability for both size-related features.

This inverse relationship between shape regularity and size variability suggests that geometric discontinuities influence melt pool formation in complex ways. The 3mm holes appear to stabilize both melt pool shape (higher regularity) and size (lower variability), while normal specimens allow greater variation in melt pool dimensions despite maintaining less regular shapes. These complementary vision features provide surface-level characterization that, when combined with the subsurface information from AE, enables comprehensive geometric variation detection.

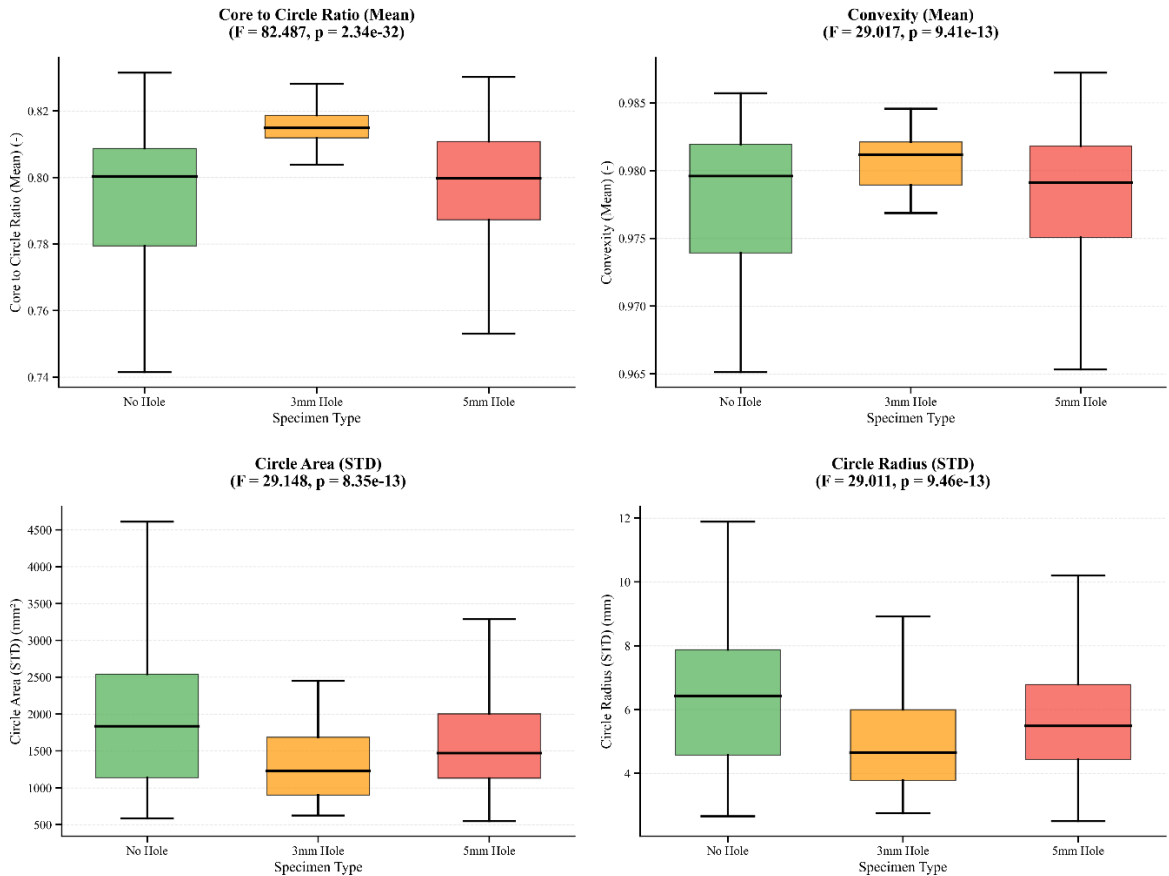

Figure 9 Box plots of discriminative vision-based features for different geometric conditions

Similar to the AE analysis, examining the temporal evolution of vision features provides insights into how melt pool characteristics develop throughout the build process. Figure 10 presents the layer-wise progression of selected vision features, revealing patterns distinct from those observed in acoustic signatures.

The shape regularity metrics—core-to-circle ratio and convexity—show that 3mm holes maintain the highest values throughout all layers, indicating consistently more regular and circular melt pools. Normal specimens and 5mm holes follow more variable trajectories, with notable crossovers occurring at different layers for each metric.

The size variability metrics—circle radius STD and circle area STD—exhibit highly correlated patterns as expected from their geometric relationship. Normal specimens consistently maintain the highest variability across all layers for both metrics. The 3mm holes show the lowest initial variability at layer 2, then increase progressively but remain below the other conditions. The 5mm holes demonstrate intermediate values throughout, maintaining clear separation from 3mm holes particularly at layers 3 and 4, before all conditions show some convergence by layer 5.

This contrast between shape regularity metrics (where 3mm holes show highest values) and size variability metrics (where normal specimens show highest values) reveals that geometric discontinuities influence melt pool formation in opposing ways—enhancing shape consistency while potentially reducing size variability. The complementary temporal patterns from vision monitoring, combined with the acoustic signatures analyzed previously, provide comprehensive characterization of how geometric variations influence the DED process across multiple layers.

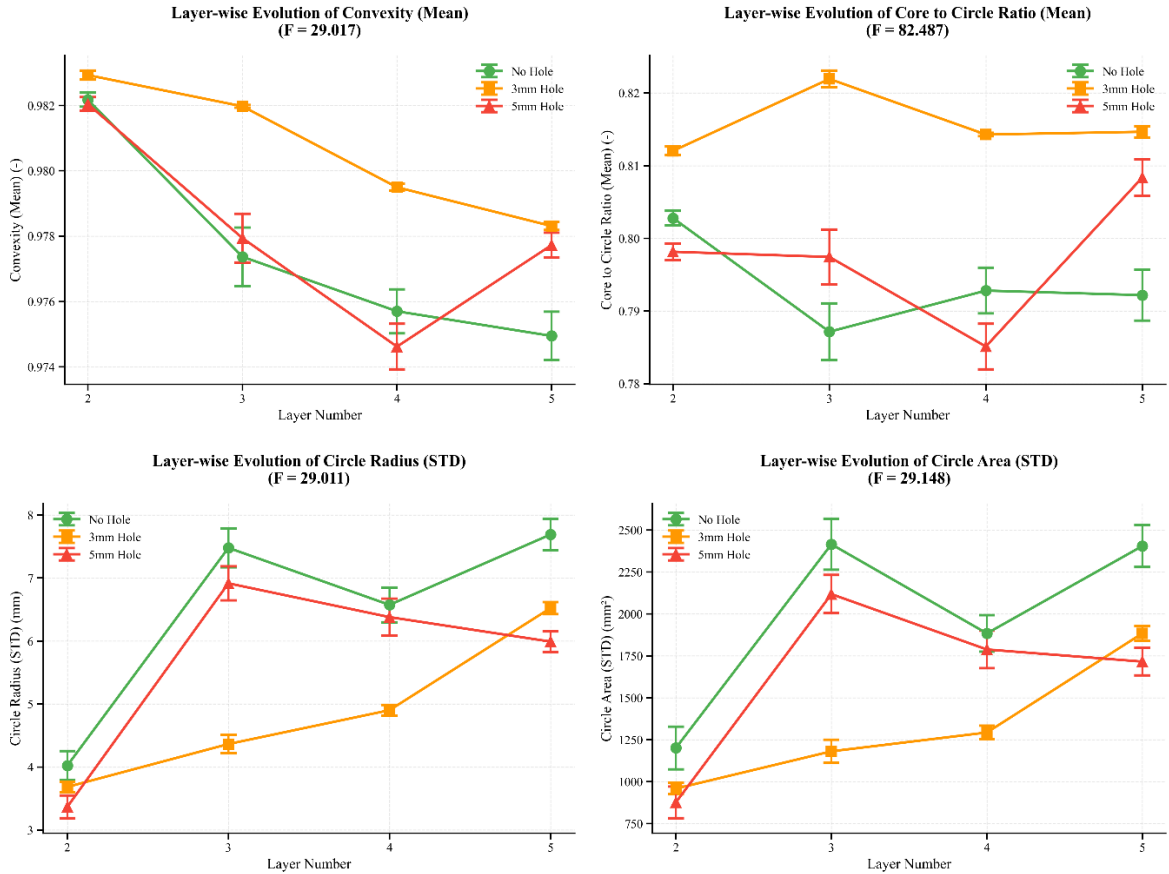

Figure 10 Layer-wise evolution of discriminative AE features for different geometric conditions

## 6.2. Multimodal machine learning results

The feature analysis in Section 6.1 established that AE and camera monitoring capture complementary aspects of the deposition process. To quantify the benefits of multimodal integration, the six machine learning architectures were trained and evaluated using the selected features from both sensing modalities.

Table 3 presents comprehensive performance metrics across all model-modality combinations. The evaluation employed the train-test split (85/15) and augmented dataset described in Sections 4.3 and 5.3, ensuring consistent comparison conditions.

Neural networks achieved the highest multimodal performance with 94.4% accuracy, 93.7% precision, and 98.3% recall. This performance approaches the stated objective of >95% accuracy for geometric variation detection. The high recall indicates exceptional sensitivity to geometric variation presence—critical for quality assurance applications where missed variations pose greater risks than false alarms.

Table 3 Classification performance metrics for different machine learning algorithms across single-modality and multimodal configurations.

| Classifier | Modality | Accuracy | Precision | Recall | F1-Score | AUC-ROC |
|---|---|---|---|---|---|---|
| **Neural Network** | **AE Only** | **0.878** | **0.889** | 0.933 | **0.911** | 0.865 |
| Neural Network | Camera Only | 0.867 | 0.853 | 0.967 | 0.906 | 0.912 |
| **Neural Network** | **Multimodal** | **0.944** | **0.937** | **0.983** | **0.959** | **0.968** |
| SVM | AE Only | 0.856 | 0.873 | 0.917 | 0.894 | 0.865 |
| SVM | Camera Only | 0.9 | 0.892 | 0.967 | 0.928 | 0.866 |
| SVM | Multimodal | 0.9 | 0.881 | 0.983 | 0.929 | 0.866 |
| **Random Forest** | **AE Only** | **0.878** | **0.889** | **0.933** | **0.911** | 0.834 |
| Random Forest | Camera Only | 0.889 | 0.879 | 0.967 | 0.921 | 0.926 |
| Random Forest | Multimodal | 0.922 | 0.921 | 0.967 | 0.943 | 0.932 |
| Gradient Boosting | AE Only | 0.867 | 0.887 | 0.917 | 0.902 | 0.887 |
| **Gradient Boosting** | **Camera Only** | **0.911** | **0.894** | **0.983** | **0.937** | 0.863 |
| Gradient Boosting | Multimodal | 0.922 | 0.921 | 0.967 | 0.943 | 0.969 |
| Logistic Regression | AE Only | 0.844 | 0.883 | 0.883 | 0.883 | 0.878 |
| Logistic Regression | Camera Only | 0.9 | 0.892 | 0.967 | 0.928 | 0.871 |
| Logistic Regression | Multimodal | 0.933 | 0.922 | 0.983 | 0.952 | 0.902 |
| XGBoost | AE Only | 0.833 | 0.8 | 1 | 0.889 | 0.802 |
| XGBoost | Camera Only | 0.856 | 0.831 | 0.983 | 0.901 | 0.836 |
| XGBoost | Multimodal | 0.933 | 0.922 | 0.983 | 0.952 | 0.936 |

Figure 11 visualizes the systematic pattern of multimodal superiority across all classifiers. Every architecture achieved its best performance with multimodal inputs, though improvement magnitudes varied considerably. The right panel quantifies these improvements relative to average single-modality performance.

Neural networks demonstrated balanced gains of 7.6% over AE-only and 9.0% over camera-only configurations, achieving an 8.3% average improvement. This balanced enhancement indicates effective integration of both information sources rather than dominance by a single modality.

Tree-based ensemble methods showed particularly strong multimodal benefits. XGBoost exhibited the most dramatic improvement at 10.5% over average single-modality performance, rising from 83.3% (AE-only) to 93.3% (multimodal)—transforming a mediocre classifier into a highly competitive system. Random Forest improved by 4.4%, while Gradient Boosting gained 3.8%. These substantial gains suggest that recursive partitioning naturally captures complex interactions between acoustic and visual features.

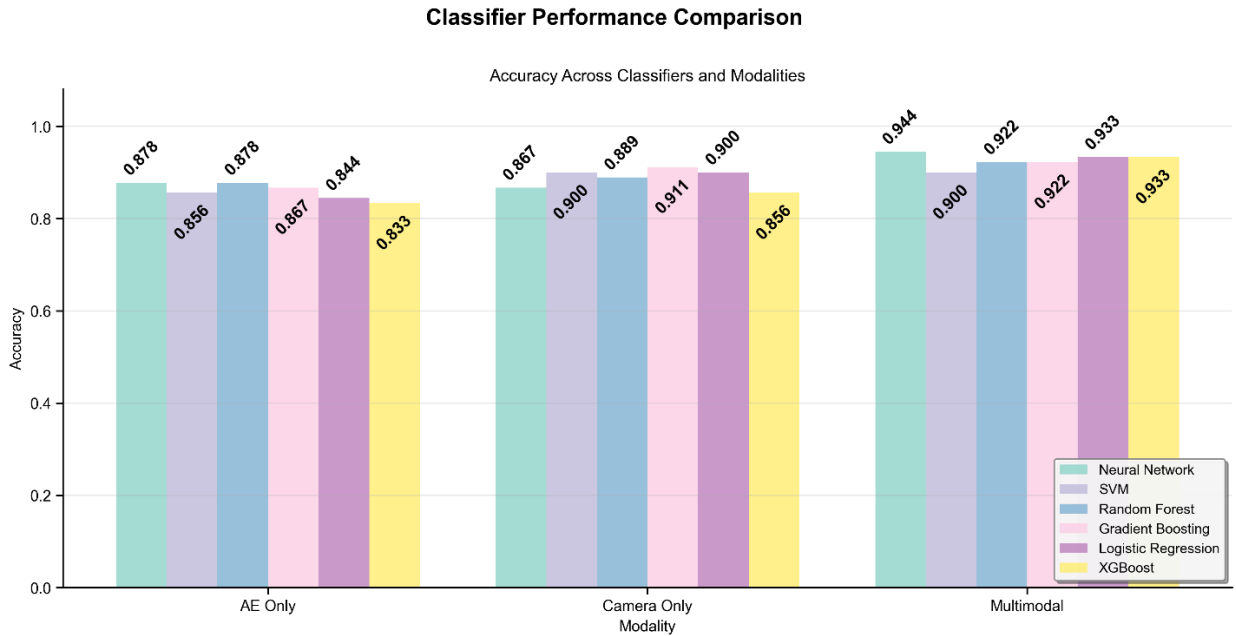

Figure 11 Comprehensive classifier performance comparison: (a) Accuracy across modalities (b) Multimodal improvement percentages

Figure 12 provides a detailed breakdown of improvements relative to each single modality. XGBoost shows 12.0% improvement over AE-only and 9.1% over camera-only, confirming its exceptional ability to leverage multimodal data. In contrast, SVM achieved only 2.5% average improvement, with multimodal performance (90.0%) matching camera-only results. This plateau indicates that kernel transformation may not effectively exploit cross-modal relationships.

Interestingly, even logistic regression achieved meaningful gains (7.0% average), demonstrating that fusion benefits exist at the basic feature level without requiring sophisticated learning mechanisms. This universal improvement pattern across fundamentally different learning paradigms confirms that performance gains arise from inherent sensor complementarity.

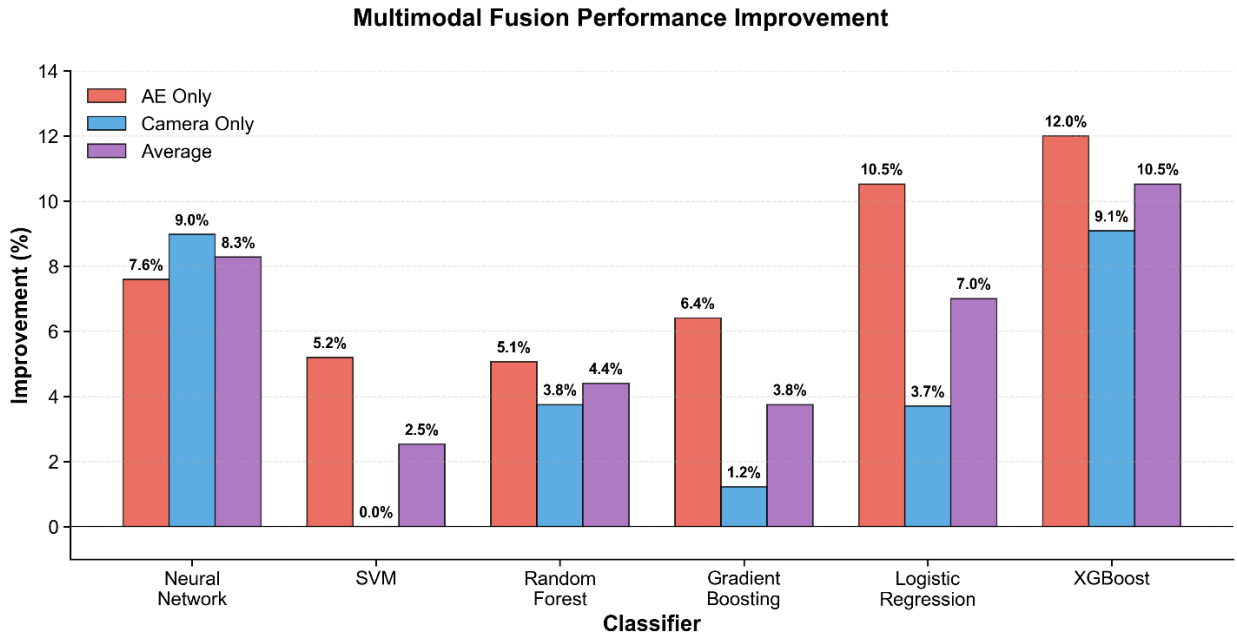

Figure 12 Performance improvement of multimodal fusion relative to single-modality approaches across different classifiers

Confusion matrix analysis provides deeper insights into error reduction mechanisms. Figure 13 reveals distinct error patterns for each configuration. AE-only classification produced 7 false positives (normal classified as defect), suggesting oversensitivity to process variations. Camera-only monitoring generated 10 false negatives (defects missed), failing to detect actual defects—particularly problematic for quality control.

The multimodal neural network dramatically reduced total errors to just 5 cases: 4 false positives and only 1 false negative. This represents a 64% error reduction compared to AE-only and 50% reduction compared to camera-only. The single false negative occurred for a 3mm hole, where early-stage signatures may not be fully developed. The error reduction pattern demonstrates that fusion effectively compensates for individual sensor limitations—AE provides high sensitivity while vision monitoring adds geometric specificity.

Performance consistency across geometric variation sizes provides additional validation. The multimodal approach correctly classified 100% of 5mm holes across all layers, reflecting strong signatures in both modalities. For the more challenging 3mm holes, multimodal fusion achieved 96.7% accuracy compared to 90.0% for AE-only and 93.3% for camera-only. This improvement for subtle defects demonstrates particular value when individual signatures approach detection thresholds.

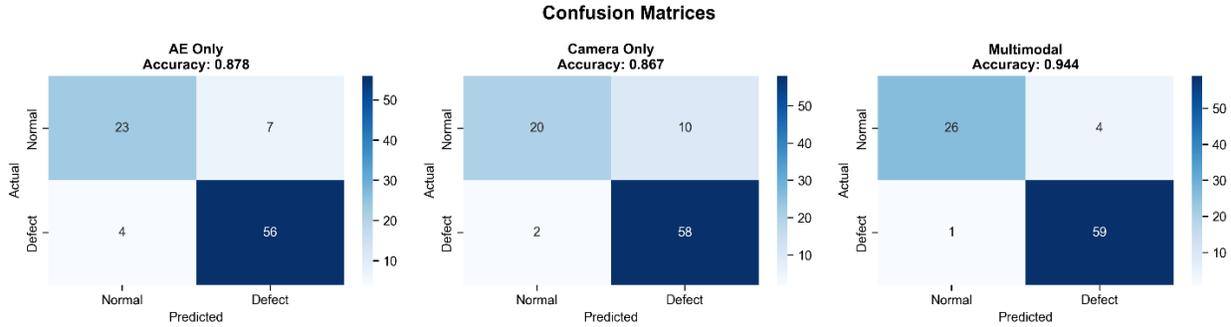

Figure 13 Confusion matrices for neural network classification across three modalities

These results establish that multimodal monitoring successfully achieves the research objective of >90% accuracy for geometric variation detection, with the neural network approaching 94.4%. The consistent improvements across diverse algorithms, coupled with dramatic error reduction and enhanced performance for challenging defects, validate the investment in dual-sensor implementation for critical quality assurance in DED manufacturing.

# 7. Conclusions

This work successfully developed and validated a multimodal monitoring system that integrates contact-based acoustic emission (AE) sensing with coaxial camera for the classification of geometric variations in laser directed energy deposition (DED). The key achievement of this work is the demonstration that the integrated multimodal approach can achieve a high classification accuracy of over 94%. This performance significantly surpasses that of either using single-modality system i.e., AE (87.8%) and camera (86.7%) operating alone, quantitatively confirming the value of fusing complementary sensor data for comprehensive process understanding.

The layer-wise analysis enables the detection of geometric variations from as early as the second layer of deposition, establishing a foundation for progressive quality assessment during the build process in DED. Furthermore, this study successfully identified distinct sets of discriminative features from both the acoustic data, which captures structural vibration dynamics, and the visual data, which reflects surface melt pool morphology. These findings confirm that the two sensing modalities capture different yet complementary physical aspects of the complex deposition process, which resulted in accurate geometric classification.

In summary, the successful demonstration of this multimodal monitoring framework establishes a validated methodology for advancing quality assurance in DED manufacturing. As this advanced manufacturing technology continues to expand into critical applications, robust monitoring systems that fuse multiple sensing modalities with intelligent analytics will be essential for ensuring part quality and process reliability. While this study used prespecified geometric variations, future work can extend these capabilities to address stochastic, manufacturing-induced defects.

## Declaration of competing interest

The authors declare that they have no known competing financial interests or personal relationships that could have appeared to influence the work reported in this paper.

## Acknowledgements

This research did not receive any specific grant from funding agencies in the public, commercial, or not-for-profit sectors.